\documentclass[11pt,epsf]{article}
\usepackage{amsmath}
\usepackage{amsfonts}
\usepackage{amssymb}
\usepackage{graphicx}
\usepackage[stable]{footmisc}

\newcommand {\dfn} {\stackrel{\Delta} {=}}
\newcommand {\exe} {\stackrel{\cdot} {=}}
\newcommand {\lexe} {\stackrel{\cdot} {\le}}

\newcommand {\br} {\mbox{\boldmath $r$}}

\newcommand {\bx} {\mbox{\boldmath $x$}}
\newcommand {\by} {\mbox{\boldmath $y$}}

\newcommand {\bE} {\mbox{\boldmath $E$}}

\newcommand {\bX} {\mbox{\boldmath $X$}}
\newcommand {\bY} {\mbox{\boldmath $Y$}}

\newcommand{\calC}{{\cal C}}

\newcommand{\calE}{{\cal E}}

\newcommand{\calI}{{\cal I}}

\newcommand{\calT}{{\cal T}}

\newcommand{\calX}{{\cal X}}
\newcommand{\calY}{{\cal Y}}

\begin{document}
\thispagestyle{empty}
\title{Erasure/List Exponents for Slepian--Wolf Decoding\thanks{
This research was supported by the Israeli Science Foundation (ISF) grant no.\
412/12.}}
\author{Neri Merhav}
\maketitle

\begin{center}
Department of Electrical Engineering \\
Technion - Israel Institute of Technology \\
Technion City, Haifa 32000, ISRAEL \\
E--mail: {\tt merhav@ee.technion.ac.il}\\
\end{center}
\vspace{1.5\baselineskip}
\setlength{\baselineskip}{1.5\baselineskip}

\begin{abstract}
We analyze random coding error exponents associated with erasure/list Slepian--Wolf
decoding using two different methods and then compare the resulting bounds. 
The first method follows the well known techniques
of Gallager and Forney and the second method is based on a technique of distance enumeration,
or more generally, type class
enumeration, which is rooted in the statistical mechanics of a disordered system
that is related to the random energy model (REM).
The second method is guaranteed to yield exponent functions which are at least
as tight as those of the first method, and it is demonstrated that for certain
combinations of coding rates and thresholds, the bounds of the second method are
strictly tighter than those of the first method, by an arbitrarily large
factor. In fact, the second method may even yield an
infinite exponent at regions where the first method gives finite values.
We also discuss the option of variable--rate Slepian--Wolf encoding
and demonstrate how it can improve on the resulting exponents.\\

\end{abstract}

\clearpage
\section{Introduction}

The celebrated paper by Slepian and Wolf \cite{SW73} has ignited
a long--lasting, intensive
research activity on separate source coding and
joint decoding of correlated
sources, during the last four decades. 
Besides its extensions in many
directions, some of the more recent studies have been devoted to further
refinements of performance
analysis, such as exponential bounds on the decoding error
probability. In particular, Gallager \cite{Gallager76} derived a lower bound
on the achievable random coding error exponent pertaining to 
random binning (henceforth, random binning exponent), using a technique that is
very similar to
that of his derivation of the ordinary random coding error exponent
\cite[Sections 5.5--5.6]{Gallager68}. This random binning exponent
was later shown by Csisz\'ar, K\"orner
and Marton \cite{CK80}, \cite{CKM77} to be universally achievable.
The work of Csisz\'ar and K\"orner \cite{CK81} is
about a universally achievable error exponent using linear codes as
well as a non--universal, expurgated exponent which is improved at high rates.
More recently, Csisz\'ar \cite{Csiszar82}
and Oohama and Han \cite{OH94} have derived error exponents for the more
general setting of coded side information.
For large rates at one of the encoders, Kelly and Wagner \cite{KW11} 
improved upon these results, but they did not consider the general case.

Since Slepian--Wolf decoding is essentially an instance of channel decoding,
we find it natural to examine its performance also in the framework of generalized
channel decoders, that is, decoders with an erasure/list option. 
Accordingly, this paper is about the analysis of random binning exponents
associated with generalized decoders.
It should be pointed
out that error exponents for list decoders of the Slepian--Wolf encoders were already analyzed
in \cite{DM07}, but in that work, it was assumed that the list size is fixed
(independent of the block length) and deterministic. In this paper, on the
hand, we analyze achievable trade-offs between random binning exponents
associated with erasure/list decoders in the framework similar to that of
Forney \cite{Forney68}. This means, among other things, that the erasure and
list options are treated jointly, on the same footing, using an optimum
decision rule of a common form, 
and that in the list option, the list
size is a random variable whose typical value might be exponentially large in
the block length. The erasure option allows the decoder not to decode 
when the confidence level is not 
satisfactory. It can be motivated, for example, by the possibility of 
generating a rate-less Slepian--Wolf code (see also \cite{EY05}), 
provided that there is at least some
minimum amount of feedback.

We analyze random binning error exponents associated with erasure/list
Slepian--Wolf
decoding using two different methods and then compare the resulting bounds.
The first method follows the well known techniques
of Gallager \cite{Gallager68} and Forney \cite{Forney68}, whereas 
the second method is based on a technique of
distance enumeration,
or more generally, on type class
enumeration. This method has already been used in previous work (see
\cite[Chapters 6--7]{Merhav09} and references therein) and proved useful in
obtaining bounds on error exponents which are always at least as
tight\footnote{It should be pointed out that in \cite{SBM11}, another
version of the type class enumeration method, which is guaranteed to yield 
the exact random coding exponents, was developed. This method, however, is
much more difficult to implement and it gives
very complicated expressions.}
(and in
many cases, strictly tighter)
than those obtained in the traditional methods of the information
theory literature. This technique 
is rooted in the statistical mechanics of certain models of disordered
magnetic materials. While in the case of ordinary random coding, the parallel
statistical--mechanical model is
the random energy model (REM) \cite[Chapters \ 5--6]{MM09}, \cite[Chapters
6--7]{Merhav09}, 
here, since random binning is considered, the
parallel statistical--mechanical model is slightly different, but related. We
will refer to this model as 
the {\it random dilution model} (RDM) for reasons that will become apparent in
the sequel. 

As mentioned in the previous paragraph, 
the type class enumeration method is guaranteed to yield an exponent function
which is at least
as tight as that of the classical method. But it is also demonstrated that for certain
combinations of coding rates and thresholds of the erasure/list decoder, 
the exponent of the type class enumeration method is
strictly tighter than that of the ordinary method. In fact, the gap between
them (i.e., their ratio) 
can be arbitrarily large, and even strictly infinite. 
In other words, for a small enough threshold
(pertaining to list decoding), the former exponent can be infinite while the latter is
finite. 

While the above described study is carried out for fixed--rate Slepian--Wolf
encoding, we also demonstrate how variable--rate encoding (with a certain
structure) can strictly improve
on the random binning exponents. This is shown in the context of the exponents
derived using the Forney/Gallager method, but a similar generalization can be
carried out using the other method.

The outline of the paper is as follows. In Section 2, we provide notation
conventions and define the objectives of the paper more formally.
In Section 3, we derive the random binning exponents using the Forney/Gallager
method, and in Section 4, we extend this analysis to allow variable rate coding.
Finally, in Section 5, after a short background on the relevant statistical--mechanical
model (Subsection 5.1),
we use the type class enumeration technique, first in the binary case
(Subsection 5.2), then compare
the resulting exponents to those of Section 3 (Subsection 5.3), and finally,
generalize the analysis to a general pair of correlated finite alphabet memoryless sources
(Subsection 5.4).

\section{Notation Conventions, Problem Formulation and Background}

\subsection{Notation Conventions}

Throughout the paper, random variables will be denoted by capital
letters, specific values they may take will be denoted by the
corresponding lower case letters, and their alphabets
will be denoted by calligraphic letters. Random
vectors and their realizations will be denoted,
respectively, by capital letters and the corresponding lower case letters,
both in the bold face font. Their alphabets will be superscripted by their dimensions. For
example, the random vector $\bX=(X_1,\ldots,X_n)$, ($n$ -- positive
integer) may take a specific vector value $\bx=(x_1,\ldots,x_n)$
in $\calX^n$, the $n$--th order Cartesian power of $\calX$, which is
the alphabet of each component of this vector. 

For a given vector $\bx$, let $\hat{P}_{\bx}$ denote the empirical
distribution, that is, the vector $\{\hat{P}_{\bx}(x),~x\in\calX\}$, where
$\hat{P}_{\bx}(x)$ is the relative frequency of the letter $x$ in the vector
$\bx$. Let
$\calT(\bx)$ denote its type class of $\bx$, namely, the set
$\{\bx':~\hat{P}_{\bx'}=\hat{P}_{\bx}\}$. The empirical entropy associated
with $\bx$, denoted $\hat{H}_{\bx}(X)$, is the entropy associated with the
empirical distribution $\hat{P}_{\bx}$. Similarly, for a pair of vectors
$(\bx,\by)$, the empirical joint distribution $\hat{P}_{\bx\by}$
is the matrix $\{\hat{P}_{\bx\by}(x,y),~x\in\calX,~y\in\calY\}$
of relative frequencies of symbol pairs $\{(x,y)\}$. The conditional type
class $\calT(\bx|\by)$ is the set $\{\bx':~\hat{P}_{\bx'\by}=\hat{P}_{\bx\by}\}$. 
The empirical conditional entropy of $\bx$ given $\by$, denoted
$\hat{H}_{\bx\by}(X|Y)$, is the conditional entropy of $X$
given $Y$, associated with the joint empirical distribution
$\{\hat{P}_{\bx\by}(x,y)\}$.

The expectation operator will be denoted by $\bE\{\cdot\}$.
Logarithms and exponents will be understood to be taken to the natural base
unless specified otherwise.
The indicator function will be denoted by $\calI(\cdot)$. The notation
function $[t]_+$ will be defined as $\max\{t,0\}$. For two positive sequences,
$\{a_n\}$ and $\{b_n\}$, the notation $a_n\exe b_n$ will mean asymptotic
equivalence in the exponential scale, that is,
$\lim_{n\to\infty}\frac{1}{n}\log(\frac{a_n}{b_n})=0$. 
Similarly, $a_n\lexe b_n$ will mean 
$\limsup_{n\to\infty}\frac{1}{n}\log(\frac{a_n}{b_n})\le 0$, and so on.

\subsection{Problem Formulation and Background}

Let $\{(X_i,Y_i)\}_{i=1}^n$ be $n$ independent copies of a random vector
$(X,Y)$, distributed according to a given probability mass function 
$P(x,y)$, where $x$ and $y$ take on values
in finite alphabets, $\calX$ and $\calY$, respectively. The source vector
$\bx=(x_1,\ldots,x_n)$, which is a generic realization of
$\bX=(X_1,\ldots,X_n)$,
is compressed at the encoder by random binning, that
is, each $n$--tuple $\bx\in\calX^n$ is randomly 
and independently assigned to one out of $M=e^{nR}$ bins,
where $R$ is the coding rate in nats per symbol.
Given a realization of the random partitioning into bins (revealed to both the encoder and
the decoder), let
$f:\calX^n\to\{0,1,\ldots,M-1\}$ denote the encoding function, i.e., $z=f(\bx)$
is the encoder output. 
Accordingly, the inverse image of $z$, defined as
$f^{-1}(z)=\{\bx:~f(\bx)=z\}$, is the bin of all source vectors mapped by the
encoder into
$z$. The decoder has access to $z$ and to $\by=(y_1,\ldots,y_n)$, which is a realization
of $\bY=(Y_1,\ldots,Y_n)$, namely, the side information at the decoder.

Following \cite{Forney68}, we consider a decoder with an
erasure/list option,
defined as follows. Let $P(\bx,\by)=\prod_{i=1}^n P(x_i,y_i)$
denote the probability of the event $\{\bX=\bx,~\bY=\by\}$ and let 
$T$ be a given real valued parameter. 
The decoding rule is as follows. For every $\hat{\bx}\in f^{-1}(z)$, if
\begin{equation}
\label{passthreshold}
\frac{P(\hat{\bx},\by)}{\sum_{\bx'\in f^{-1}(z)\setminus\{\hat{\bx}\}} P(\bx',\by)}\ge
e^{nT},
\end{equation}
then $\hat{\bx}$ is referred to as a {\it candidate}. If there are no
candidates, an
erasure is declared, namely, the decoder acts in its erasure mode. 
If there is exactly one candidate,
$\hat{\bx}$, then this is the estimate that the decoder produces, just
like in ordinary decoding.
Finally, if there is more than one candidate, then the decoder
operates in the list mode and it outputs the list of all candidates.
Obviously, for $T\ge 0$, the list can contain at most one candidate. The list
may contain two candidates or more only for sufficiently small negative values of $T$.

Forney \cite{Forney68} used the Neymann--Pearson lemma, in an analogous
channel coding setting, to show that 
the above rule simultaneously gives rise to: (i) an optimum
trade-off between the probability of erasure and the probability of decoding error, in
the erasure mode, and (ii) an optimum trade-off between the probability of list
error and the expected number of
incorrect candidates on the list, in the list mode.
Our goal, in this paper, is to assess the exponential rates associated with these trade-offs.

\section{Error Exponent Analysis Based on the Gallager/Forney Method}

Similarly as in \cite{Forney68}, we define the event 
$\calE_1$ as the event that the correct source
vector $\bx$ is not a candidate, that is, 
\begin{equation}
\frac{P(\bx,\by)}
{\sum_{\bx'\in f^{-1}(z)\setminus\{\bx\}}P(\bx',\by)}
<e^{nT}.
\end{equation}
We next derive a lower bound on the exponential rate $E_1(R,T)$ of 
the average probability of $\calE_1$, where the averaging is with
respect to (w.r.t.) the ensemble of random binnings. The other exponent,
$E_2(R,T)$ (of decoding error in the erasure option, or the expected list size in list
option) will then be given by $E_2(R,T)=E_1(R,T)+T$, similarly as in
\cite{Forney68}. We now have the following chain of inequalities for any $s\ge 0$:
\begin{eqnarray}
\label{beginning}
\mbox{Pr}\{\calE_1\}&=&\sum_{\bx,\by}P(\bx,\by)\calI\left\{\frac{e^{nT}\sum_{\bx'\ne\bx}
P(\bx',\by)\calI[f(\bx')=f(\bx)]}{P(\bx,\by)}> 1\right\}\nonumber\\
&\le&\sum_{\bx,\by}P(\bx,\by)\left[\frac{e^{nT}\sum_{\bx'\ne\bx}
P(\bx',\by)\calI[f(\bx')=f(\bx)]}{P(\bx,\by)}\right]^s\nonumber\\
&=&e^{nsT}\sum_{\bx,\by}P^{1-s}(\bx,\by)\left[\sum_{\bx'\ne\bx}
P(\bx',\by)\calI[f(\bx')=f(\bx)]\right]^s.
\end{eqnarray}
Now, let $\rho\ge s$ be another parameter. Then,
\begin{eqnarray}
\mbox{Pr}\{\calE_1\}&\le&e^{nsT}\sum_{\bx,\by}P^{1-s}(\bx,\by)\left(\left[\sum_{\bx'\ne\bx}
P(\bx',\by)\calI[f(\bx')=f(\bx)]\right]^{s/\rho}\right)^\rho\\
&\le&e^{nsT}\sum_{\bx,\by}P^{1-s}(\bx,\by)\left(\sum_{\bx'\ne\bx}
P^{s/\rho}(\bx',\by)\calI[f(\bx')=f(\bx)]\right)^\rho.
\end{eqnarray}
where we have used the inequality $(\sum_i a_i)^t\le\sum_i a_i^t$ for
$t\in[0,1]$.
Taking now the expectation w.r.t.\ the randomness of the binning, and assuming
that $\rho\le 1$, we get
\begin{eqnarray}
\overline{\mbox{Pr}\{\calE_1\}}&\le&
e^{nsT}\sum_{\bx,\by}P^{1-s}(\bx,\by)\bE\left\{\left(\sum_{\bx'\ne\bx}
P^{s/\rho}(\bx',\by)\calI[f(\bx')=f(\bx)]\right)^\rho\right\}\\
&\le&e^{nsT}\sum_{\bx,\by}P^{1-s}(\bx,\by)\left(\sum_{\bx'\ne\bx}
P^{s/\rho}(\bx',\by)\bE\{\calI[f(\bx')=f(\bx)]\}\right)^\rho\\
&=&e^{nsT}\sum_{\bx,\by}P^{1-s}(\bx,\by)\left(\sum_{\bx'\ne\bx}
P^{s/\rho}(\bx',\by)e^{-nR}\right)^\rho\\
&=&e^{-n(\rho R-sT)}\sum_{\bx,\by}P^{1-s}(\bx,\by)\left(\sum_{\bx'\ne\bx}
P^{s/\rho}(\bx',\by)\right)^\rho\\
&=&e^{-n(\rho R-sT)}\sum_{\by}P(\by)\sum_{\bx}P^{1-s}(\bx|\by)\left(\sum_{\bx'\ne\bx}
P^{s/\rho}(\bx'|\by)\right)^\rho\\
&\le&e^{-n(\rho
R-sT)}\left[\sum_{y\in\calY}P(y)\sum_{x\in\calX}P^{1-s}(x|y)\left(\sum_{x'\in\calX}
P^{s/\rho}(x'|y)\right)^\rho\right]^n.
\end{eqnarray}
Thus, after optimization over $\rho$ and $s$, subject to the constraints
$0\le s\le\rho\le 1$, we obtain
\begin{equation}
\overline{\mbox{Pr}\{\calE\}}\le e^{-nE_1(R,T)}
\end{equation}
where
\begin{equation}
E_1(R,T)=\sup_{0\le s\le\rho\le 1}[E_0(\rho,s)+\rho R-sT]
\end{equation}
with
\begin{equation}
E_0(\rho,s)=-\ln\left[\sum_{y\in\calY}P(y)\sum_{x\in\calX}P^{1-s}(x|y)\left(\sum_{x'\in\calX}
P^{s/\rho}(x'|y)\right)^\rho\right].
\end{equation}
A few elementary properties of the function
$E_1(R,T)$ are the following.
\begin{enumerate}
\item $E_1(R,T)$ is jointly convex in both arguments. This follows directly from the fact
that it is given by the supremum over a family of affine functions in $(R,T)$.
Clearly, $E_1(R,T)$ is increasing in $R$ and decreasing in $T$.
\item At $T=0$, the optimum $s$ is $\rho/(1+\rho)$, similarly as in
\cite{Forney68} and
\cite{Gallager76}. Thus, as observed in \cite{Forney68}, here too, the case
$T=0$ is essentially equivalent (in terms of error exponents) to ordinary
decoding, although operationally, there still might
be erasures in this case.
\item For a given $T$, the infimum of $R$ such that $E_1(R,T)>0$ is
\begin{equation}
R_{\min}(T)=\inf_{0\le s\le \rho\le 1}\frac{sT-E_0(\rho,s)}{\rho},
\end{equation}
which is a concave increasing function. At $T=0$,
$$R_{\min}(0)=-
\sup_{0\le \rho\le 1}\frac{E_0\left(\rho,\frac{\rho}{1+\rho}\right)}{\rho}=
-\lim_{\rho\to 0}\frac{E_0\left(\rho,\frac{\rho}{1+\rho}\right)}{\rho}=
-\frac{\partial}{\partial\rho}E_0\left(\rho,\frac{\rho}{1+\rho}\right)\bigg|_{\rho=0}=
H(X|Y).$$ 
\item For a given $R$, the supremum of $T$ such that $E_1(R,T)>0$ is
\begin{equation}
T_{\max}(R)=\sup_{0\le s\le \rho\le 1}\frac{\rho R+E_0(\rho,s)}{s},
\end{equation}
which is a convex increasing function, the inverse of $R_{\min}(T)$.
\end{enumerate}
Additional properties can be found similarly as in \cite{Forney68}, but we
will not delve into them here.

\section{Extension to Variable--Rate Slepian--Wolf Coding}

A possible extension of the above error exponent analysis allows variable rate coding. 
In this section, we demonstrate how the flexibility of variable--rate coding
can improve the error exponents. 

Consider an encoder that first sends a relatively 
short header that encodes the type class of $\bx$
(using a logarithmic number of bits), and then a description of 
$\bx$ within its type class, using a random bin $z=f(\bx)$ in the range
$\{0,1,\ldots,\exp[nR(\bx)]-1\}$, where $R(\bx)> 0$ depends on $\bx$ only via the
type class of $\bx$. The bin $z$ for every $\bx$ in its type class is selected independently at
random with a uniform probability distribution $P(z)=e^{-nR(\bx)}$. The average
coding rate would be, of course, $R=\bE\{R(\bX)\}$ (neglecting the rate of the
header). For example, consider an additive rate
function\footnote{The reason for choosing a rate function with this simple
structure is that it allows to easily generalize the analysis in the
Gallager/Forney style and obtain single--letter expressions
without recourse to the method of types. More general rate functions, that
depend on the type class of $\bx$ in an arbitrary manner, are still manageable, but
require the method of types.}
$R(\bx)=\frac{1}{n}\sum_{i=1}^n r(x_i)$. Thus,
$R=\bE\{r(X)\}=\sum_{x\in\calX}P(x)r(x)$. Extending the above error exponent
analysis, one readily obtains\footnote{Observe that here
$\mbox{Pr}\{f(\bx')=f(\bx)\}=e^{-nR(\bx')}$ 
whenever $e^{nR(\bx')}< f(\bx)$ and
$\mbox{Pr}\{f(\bx')=f(\bx)\}=0$ 
elsewhere, thus $\mbox{Pr}\{f(\bx')=f(\bx)\}\le e^{-nR(\bx')}$ everywhere.}
\begin{equation}
\tilde{E}_1(R,T)=\sup_{0\le s\le\rho\le 1}\sup_{\{\br:~\bE\{r(X)\}\le
R,~r(x)> 0~\forall~x\in\calX\}}[\tilde{E}_0(\rho,s)-sT],
\end{equation}
where $\br\dfn\{r(x),~x\in\calX\}$ 
and where $\tilde{E}_0(\rho,s)$ is defined as
\begin{equation}
\tilde{E}_0(\rho,s)=-\ln\left[\sum_{y\in\calY}P(y)
\sum_{x\in\calX}P^{1-s}(x|y)\left(\sum_{x'\in\calX}
P^{s/\rho}(x'|y)e^{-r(x')}\right)^\rho\right].
\end{equation}
It is interesting to find the optimum rate assignment
$\br=\{r(x)~x\in\calX\}$ that maximizes the exponent. Consider, for example,
the case where $R$ and $T$ are such that $E_1(R,T)$ is achieved by $\rho=1$.
Then,
\begin{eqnarray}
e^{-E_0(1,s)}&=&\sum_{y\in\calY}P(y)\sum_{x\in\calX}P^{1-s}(x|y)\sum_{x'\in\calX}
P^s(x'|y)e^{-r(x')}\\
&=&\sum_{x\in\calX}F(x)e^{-r(x)}
\end{eqnarray}
where
\begin{equation}
F(x)\dfn\sum_{y\in\calY}P(y)P^s(x|y)\sum_{x'\in\calX}P^{1-s}(x'|y).
\end{equation}
Our task now is to minimize $\sum_{x\in\calX}F(x)e^{-r(x)}$ subject to the constraints
$\sum_{x\in\calX}P(x)r(x)\le R$ and $r(x) > 0$ for all $x\in\calX$, which is a
standard convex program.
For simplicity, let us first ignore the constraints $r(x)>0$, $x\in\calX$, and assume
that the parameters of the problem are such that the resulting solution
will satisfy these positivity constraints anyway. Then, 
\begin{equation}
r^*(x)=\lambda+\ln\frac{F(x)}{P(x)},
\end{equation}
where $\lambda$ is determined by the average rate constraint, that is
\begin{eqnarray}
\lambda&=&R+\sum_{x\in\calX}P(x)\ln\frac{P(x)}{F(x)}\\
&=&R+D(P\|Q)-\ln\left[\sum_{x\in\calX}F(x)\right],
\end{eqnarray}
where 
\begin{equation}
Q(x)=\frac{F(x)}{\sum_{x'\in\calX}F(x')}.
\end{equation}
Thus,
\begin{equation}
r^*(x)=R+D(P\|Q)+\ln\frac{Q(x)}{P(x)}.
\end{equation}
We see that fixed--rate coding is optimum only if $P(x)$ happens to be
proportional to $F(x)$, namely, $P=Q$ (which is the case,
for example, when $s=1$). Upon substituting $\br^*=\{r^*(x),~x\in\calX\}$ back into the
objective function, we obtain
\begin{eqnarray}
e^{-\tilde{E}_0(1,s)}&=&\sum_{x\in\calX}F(x)\exp\{-\lambda-\ln[F(x)/P(x)]\}\\
&=&\sum_{x\in\calX}P(x)e^{-\lambda}
=e^{-\lambda},
\end{eqnarray}
and so,
\begin{eqnarray}
\tilde{E}_0(1,s)&=&\lambda\\
&=&R+D(P\|Q)-\ln\left[\sum_{x\in\calX}F(x)\right]\\
&=&R+D(P\|Q)-\ln\left[\sum_{y\in\calY}P(y)\sum_{x\in\calX}P^{1-s}(x|y)\sum_{x'\in\calX}P^s(x'|y)\right]\\
&=&E_0(1,s)+D(P\|Q).
\end{eqnarray}
The term $D(P\|Q)$ then represents the improvement we have obtained 
upon passing from fixed---rate coding to variable--rate coding with
an additive rate function. This is true for a given $s$. However, after
re--optimizing the bound over $s$, the improvement can be even larger.
When $R+D(P\|Q)+\ln[Q(x)/P(x)]$ are not all
positive, the optimum solution is given by
\begin{equation}
r^*(x)=\left[\ln\frac{Q(x)}{P(x)}+\mu\right]_+
\end{equation}
where $\mu$ is the (unique) solution to the equation
\begin{equation}
\sum_{x\in\calX}P(x)\left[\ln\frac{Q(x)}{P(x)}+\mu\right]_+=R.
\end{equation}
For $\rho < 1$, the optimization over $\br$ is
less trivial, but it can still be carried out at least numerically.

\section{Error Exponent Analysis Using Type Class Enumeration}

\subsection{A Brief Background in Statistical Mechanics}

This subsection can be skipped without essential loss of continuity, however,
we believe that before
getting into the detailed technical derivation, it would be
instructive to give a brief review of the statistical--mechanical models that
are at the basis of the type class enumeration method. 

In ordinary random coding (as opposed to random binning), the derivations of
bounds on the error probability (especially in the methods of
Gallager and Forney) are frequently associated with expressions of the form
$\sum_{\bx\in\calC} P^\beta(\by|\bx)$, where $\calC$ is (randomly selected)
codebook and $\beta > 0$ is some parameter. 
As explained in \cite[Chap.\ 6]{Merhav09}, this can be viewed, from the
statistical--mechanical perspective, as a partition function
\begin{equation}
Z(\beta)=\sum_{\bx\in\calC} e^{-\beta E(\bx,\by)},
\end{equation}
where $\beta$ plays the role of inverse temperature
and where the energy function (Hamiltonian) 
is $E(\bx,\by)=-\ln P(\by|\bx)$. Since the codewords are selected
independently at
random, then for a given $\by$, the energies $\{E(\bx,\by),~\bx\in\calC\}$ are
i.i.d.\ random variables. This is, in principle, nothing but the {\it random energy
model} (REM), a well known model in statistical mechanics of disordered magnetic
materials (spin glasses), which exhibits a phase transition: below a certain
critical temperature ($\beta > \beta_c$), the system freezes in the sense that the
partition function is exponentially dominated by a subexponential number of
configurations at the ground--state energy (zero thermodynamical entropy). This phase is called the
{\it frozen phase} or the {\it glassy phase}. 
The other phase, $\beta < \beta_c$, is called the
{\it paramagnetic phase}
(see more details in \cite[Chap.\
5]{MM09}). Accordingly, the resulting exponential error bounds associated
with random coding `inherit' this phase transition (see \cite{Merhav09}
and references therein).

In random binning the situation is somewhat different. As we have seen in
Section 3, here the bound involves an expression like
$\sum_{\bx'}P^\beta(\bx',\by)\calI[f(\bx')=f(\bx)]$. The source vectors $\{\bx'\}$
that participate in the summation 
are now deterministic, but the random ingredient is the function $f$. The
analogous statistical--mechanical model is then
encoded into the partition function
\begin{equation}
Z(\beta)=\sum_{\bx} I(\bx)\cdot e^{-\beta E(\bx,\by)},
\end{equation}
where $\{I(\bx),~\bx\in\calX^n\}$ are i.i.d.\ binary random variables, taking on values in
$\{0,1\}$, where $\mbox{Pr}\{I(\bx)=1\}=e^{-nR}$. In other words, $Z(\beta)$ is
a randomly diluted version of the full partition function $\sum_{\bx}e^{-\beta
E(\bx,\by)}$, where each configuration $\bx$ `survives' with probability
$e^{-nR}$ or is discarded with probability $1-e^{-nR}$. 
Accordingly, we refer to this model as the {\it random dilution model} (RDM).
To the best of our knowledge, such a model has not been used in statistical
mechanics thus far, but it
can be analyzed in the very same fashion, and it is easy to see that it also
exhibits a glassy phase transition (depending on $R$). In fact, the RDM can be considered as a variant of
the REM, where the configurational energies are $E(\bx,\by)+\phi(\bx)$, where $\phi(\bx)=0$
with probability $e^{-nR}$ and $\phi(\bx)=\infty$
with probability $1-e^{-nR}$. Thus, $\phi(\bx)$ can be thought of as
disordered potential function, associated with long--range interactions, with
infinite spikes that forbid
access to certain points in the configuration space.

\subsection{The Binary Case}

\begin{figure}[ht]
\hspace*{3cm}\input{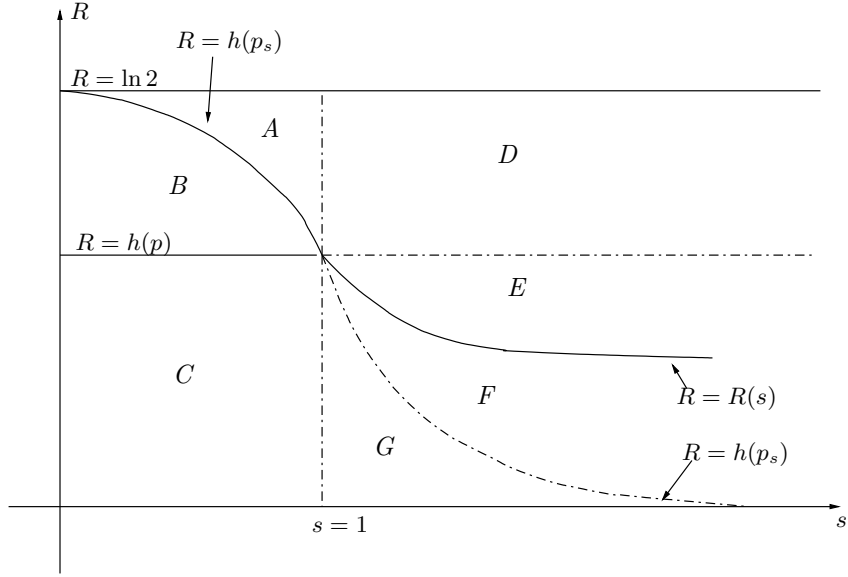}
\caption{\small Phase diagram of the function $L(R,s)$.}
\label{swepd}
\end{figure}

Let us return to the fixed--rate regime.
It is instructive to begin from the relatively simple special case where $\bX$
and $\bY$ are correlated binary symmetric sources (BSS's), that is,
\begin{equation}
P(x,y)=\left\{\begin{array}{ll}
(1-p)/2 & x=y\\
p/2 & x\ne y\end{array}\right.~~~~~~~~x,y\in\{0,1\}
\end{equation}
We begin similarly as in Section 3: Our starting point is
the same bound as in the last line of eq.\ (\ref{beginning}), specialized to
the binary case considered here, where we also take the ensemble average:
\begin{eqnarray}
\overline{\mbox{Pr}\{\calE_1\}}&\le&
e^{nsT}\sum_{\bx,\by}P^{1-s}(\bx,\by)
\cdot\bE\left\{\left[\sum_{\bx'\ne\bx}P(\bx',\by)\calI[f(\bx')=f(\bx)]\right]^s\right\}\\
&=&e^{nsT}\sum_{\by}P(\by)\left[\sum_{\bx}P^{1-s}(\bx|\by)\right]
\cdot\bE\left\{\left[\sum_{\bx'\ne\bx}P(\bx'|\by)\calI[f(\bx')=f(\bx)]\right]^s\right\}\\
&=&e^{nsT}\sum_{\by}2^{-n}\left[p^{1-s}+(1-p)^{1-s}\right]^n
\cdot\bE\left\{\left[\sum_{\bx'\ne\bx}P(\bx'|\by)\calI[f(\bx')=f(\bx)]\right]^s\right\}\\
&=&e^{nsT}\left[p^{1-s}+(1-p)^{1-s}\right]^n
\cdot\bE\left\{\left[\sum_{\bx'\ne\bx}P(\bx'|\by)\calI[f(\bx')=f(\bx)]\right]^s\right\}
\end{eqnarray}
where the last step is justified by the fact that the expectation term is
independent of $\by$, as will be seen shortly.
Now,
\begin{eqnarray}
\bE\left\{\left[\sum_{\bx'\ne\bx}P(\bx'|\by)\calI[f(\bx')=f(\bx)]\right]^s\right\}
&\exe&\sum_{\calT(\bx'|\by)}P^s(\bx'|\by)\bE\{N^s(\bx'|\bx,\by)\}\\
&=&(1-p)^{ns}\sum_{\delta}\left(\frac{p}{1-p}\right)^{ns\delta}\bE\{N^s(\bx'|\bx,\by)\}
\end{eqnarray}
where $\delta$ is the normalized Hamming distance, the summation is
over the set $\{0,1/n,2/n,\ldots,1-1/n,1\}$, and
$N(\bx'|\bx,\by)=\sum_{\bx'\in \calT(\bx|\by)}\calI[f(\bx')=f(\bx)]$.
Now, $N(\bx'|\bx,\by)$ is the sum of $|\calT(\bx|\by)|\exe
\exp\{n\hat{H}_{\bx\by}(X|Y)\}$ i.i.d.\ binary random variables 
$\{\calI[f(\bx')=f(\bx)]\}$ with $\mbox{Pr}\{f(\bx')=f(\bx)\}=e^{-nR}$.
Thus, similarly as in \cite[Sect.\ 6.3]{Merhav09}
\begin{eqnarray}
\bE\{N^s(\bx'|\bx,\by)\}&\exe&\left\{\begin{array}{ll}
\exp\{ns[h(\delta)-R]\} & h(\delta)\ge R\\
\exp\{n[h(\delta)-R]\} & h(\delta)< R \end{array}\right.\\
&=& \exp\{n(s[h(\delta)-R]-(1-s)[R-h(\delta)]_+)\}
\end{eqnarray}
and so
\begin{eqnarray}
\overline{\mbox{Pr}\{\calE_1\}}&\le&
e^{nsT}\left[p^{1-s}+(1-p)^{1-s}\right]^n(1-p)^{ns}\sum_\delta
\left(\frac{p}{1-p}\right)^{ns\delta}\times\nonumber\\
& &\exp\{n(s[h(\delta)-R]-(1-s)[R-h(\delta)]_+)\}\\
&\exe& e^{nsT}\left[p^{1-s}+(1-p)^{1-s}\right]^n(1-p)^{ns}e^{-nL(R,s)}
\end{eqnarray}
where $L(R,s)\dfn\min_{0\le \delta\le 1}L(R,s,\delta)$ with
\begin{equation}
L(R,s,\delta)\dfn s\delta\ln\frac{1-p}{p}+s[R-h(\delta)]+(1-s)[R-h(\delta)]_+.
\end{equation}
Standard optimization of $L(R,s,\delta)$ gives the following result (see
Appendix A for the details).
Define the sets (see also Fig.\ \ref{swepd})
\begin{eqnarray}
A&=&\{(s,R): 0\le s\le 1,~R>h(p_s)\}\\
B&=&\{(s,R): 0\le s\le 1,~h(p)< R\le h(p_s)\}\\
C&=&\{(s,R): 0\le s\le 1,~R\le h(p)\}\\
D&=&\{(s,R): s> 1,~R> h(p)\}\\
E&=&\{(s,R): s> 1,~R(s)< R\le h(p)\}\\
F&=&\{(s,R): s> 1,~h(p_s)< R\le R(s)\}\\
G&=&\{(s,R): s> 1,~R\le h(p_s)\}.
\end{eqnarray}
Then,
\begin{equation}
L(R,s)=\left\{\begin{array}{ll}
s[p\ln\frac{1-p}{p}+R-h(p)] & (s,R)\in C\cup F\cup G\\
sh^{-1}(R)\ln\frac{1-p}{p} & (s,R)\in B\\
sp_s\ln\frac{1-p}{p}+R-h(p_s) & (s,R)\in A\cup D\cup E
\end{array}\right.
\end{equation}
Finally, the exponent of $\overline{\mbox{Pr}\{E_1\}}$ is lower bounded by
\begin{equation}
E_1'(R,T)=\sup_{s\ge
0}\left\{L(R,s)+s\ln\frac{1}{1-p}-\ln[p^{1-s}+(1-p)^{1-s}]-sT\right\}.
\end{equation}
Equivalently, $E_1'(R,T)$ can be presented as follows:
\begin{equation}
E_1'(R,T)=\sup_{s\ge
0}E_1'(R,T,s)
\end{equation}
where
\begin{equation}
E_1'(R,T,s)=\left\{\begin{array}{ll}
s(R-T)-\ln[p^{1-s}+(1-p)^{1-s}] & (s,R)\in C\cup F\cup G\\
s[R-T+D(h^{-1}(R)\|p)]-\ln[p^{1-s}+(1-p)^{1-s}] & (s,R)\in B\\
R-sT-\ln[p^s+(1-p)^s]-\ln[p^{1-s}+(1-p)^{1-s}] & (s,R)\in A\cup D\cup E
\end{array}\right.
\end{equation}

Fig.\ 1 depicts a phase diagram of the function $L(R,s)$. 
This function inherits phase transitions associated with the 
analogous statistical--mechanical model -- the RDM. The strip
defined by $s\ge 0$ and $0\le R\le \ln 2$ is divided into seven regions,
labeled by the letters A--G as defined above. There are three main phases that are separated by solid
lines, which differ in terms of the expression of $L(R,s)$.
The phase $C\cup F\cup G$ is the phase where typical realiztions of the
random binning ensemble dominate the partition function (that is, conditional
type classes of size less
than $e^{nR}$ contain no matching bin, whereas conditional type classes of larger size have an
exponentially typical number of bin matches), phase $B$ is
the glassy phase,
and phase $A\cup D\cup E$ is the phase where the conditional small type classes dominate
the partition function (unlike in phase $C\cup F\cup G$).
A secondary partition into sub--phases (dashed lines) correspond to different
shapes of the objective function $L(R,s,\delta)$. In regions $A$, $B$, $C$ ($s\le
1$), the
derivative of the objective function has a positive jump at
$\delta=h^{-1}(R)$, and the minimizer is smaller than $h^{-1}(R)$,
equal to $h^{-1}(R)$, and larger than $h^{-1}(R)$, respectively.
In regions $D$, $E$, $F$ and $G$ ($s>1$), the derivative of
$L(R,s,\delta)$ w.r.t.\ $\delta$ has a
negative jump at $\delta=h^{-1}(R)$, In regions $E$ and $F$, this jump is from
a positive derivative to a negative derivative, meaning that
$\delta=h^{-1}(R)$ is a
(non--smooth) local maximum and there are two local minima,
one at $\delta =p < h^{-1}(p)$ and one at $\delta=p_s > h^{-1}(R)$.
In region $E$, the local minimum at $\delta=p_s$
is smaller than the local minimum at $\delta=p$ and in region $F$ it is
vice versa. In region $G$ there is only one local minimum at $\delta=p$
and in region $D$ there is only one local minimum at $\delta=p_s$.

\subsection{Comparison of the Exponents}

The expression of $E_1'(R,T)$ should be compared with $E_1(R,T)$ specialized
to the double BSS considered in Subsection 5.2, i.e.,
\begin{equation}
E_1(R,T)=\sup_{0\le s\le \rho\le 1}\left\{
\rho R-\ln[p^{1-s}+(1-p)^{1-s}]-\rho\ln[p^{s/\rho}+(1-p)^{s/\rho}]-sT\right\}.
\end{equation}
Obviously, $E_1'(R,T)\ge E_1(R,T)$ since derivation of $E_1'(R,T)$ is
guaranteed to be exponentially tight starting from (\ref{beginning}),
in contrast to the derivation of $E_1(R,T)$, which is associated with Jensen's
inequality, as well as the inequality $(\sum_ia_i)^t\le\sum_ia_i^t$, $0\le
t\le 1$, following \cite{Forney68}.

To show an extreme situation of a
strict inequality, $E_1'(R,T)> E_1(R,T)$, 
consider the case where $R> h(p)$ and $T < \ln[p/(1-p)] <
0$ (a list
option). Then,
\begin{eqnarray}
E_1'(R,T)&\ge&\lim_{s\to\infty}\left\{R-sT-
\ln\left[(1-p)^s\left(1+\left[\frac{p}{1-p}\right]^s\right)\right]-\right.\nonumber\\
& &\left.\ln\left[p^{1-s}\left(1+\left[\frac{1-p}{p}\right]^{1-s}\right)\right]\right\}\\
&=&\lim_{s\to\infty}\left\{R-sT-s\ln(1-p)-
\ln\left(1+\left[\frac{p}{1-p}\right]^s\right)-(1-s)\ln p-\right.\nonumber\\
& &\left.\ln\left(1+\left[\frac{p}{1-p}\right]^{s-1}\right)\right\}\\
&=&\lim_{s\to\infty}\left\{R-sT-s\ln(1-p)
-(1-s)\ln p\right\}\\
&=&\ln\frac{1}{p}+R+\lim_{s\to\infty}s\left[\ln\frac{p}{1-p}-T\right]\\
&=&\infty.
\end{eqnarray}
On the other hand, in this case,
\begin{eqnarray}
E_1(R,T)&\le&R+|T|+2\max_{0\le \alpha\le 1}\{-\ln[p^\alpha+(1-p)^\alpha]\}\\
&=& R+|T| < \infty.
\end{eqnarray}

Another situation, where it is relatively easy to calculate the exponents
is the limit of very weak correlation between the BSS's $X$ and $Y$ (in analogy to the
notion of a very noisy channel \cite[p.\ 147, Example 3]{Gallager68}).
Let $p=1/2-\epsilon$
for $|\epsilon| \ll 1$. 
In this case, a second order Taylor series expansion of the relevant functions
(see Appendix B for the details) yields, for $h(p)\le R\le \ln 2$ and $T=-\tau\epsilon^2$,
with $\tau > 4$ being fixed:
\begin{equation}
E_1(R,T)\le
(\tau+2)\epsilon^2,
\end{equation}
whereas
\begin{equation}
E_1'(R,T)\ge\left[\frac{\tau(\tau+8)}{16}-1\right]\epsilon^2.
\end{equation}
Now, observe that the upper bound on $E_1(R,T)$ is affine in $\tau$, whereas
the lower bound on $E_1'(R,T)$ is quadratic in $\tau$, thus the ratio
$E_1'(R,T)/E_1(R,T)$ can be made arbitrarily large for any sufficiently large
$\tau > 4$. 

In both examples, we took advantage of the fact that the range of optimization of $s$ for
$E_1'(R,T)$ includes all the positive reals, whereas for $E_1(R,T)$, it is
limited to the interval $[0,1]$ due to the combination of using of Jensen's
inequality (which requires $\rho\le 1$)
and the inequality $(\sum_ia_i)^t\le\sum_ia_i^t$ (which requires
$s\le\rho$). Note that the second example is not a special case the first one,
because in the first example, for $p=1/2-\epsilon$, $|T|>
\ln[(1-p)/p]=O(\epsilon)$, whereas in the second example, $T=
O(\epsilon^2)$.

\subsection{Extension to General Finite Alphabet Memoryless Sources}

In this subsection, we use the type class enumeration method for general
finite alphabet sources $S$ and $Y$.
Consider the expression
$$\bE\left\{\left[\sum_{\bx'\ne\bx}
P(\bx',\by)\calI[f(\bx')=f(\bx)]\right]^s\right\}$$
that appears upon taking the expectation over the last line of (\ref{beginning}). 
Then, we have
\begin{eqnarray}
&&\bE\left\{\left[\sum_{\bx'\ne\bx}
P(\bx',\by)\calI[f(\bx')=f(\bx)]\right]^s\right\}\\
&=&P^s(\by)\bE\left\{\left[\sum_{\bx'\ne\bx}
P(\bx'|\by)\calI[f(\bx')=f(\bx)]\right]^s\right\}\\
&\le&P^s(\by)\sum_{\calT(\bx'|\by)}P^s(\bx'|\by)\bE\left\{\left[\sum_{\tilde{\bx}\in
\calT(\bx'|\by)}
\calI[f(\tilde{\bx})=f(\bx)]\right]^s\right\}\\
&\dfn&P^s(\by)\sum_{\calT(\bx'|\by)}P^s(\bx'|\by)\bE\left\{N^s(\bx'|\bx,\by)\right\}
\end{eqnarray}
where $N(\bx'|\bx,\by)$ is the (random) number of $\{\tilde{\bx}\}$ in
$\calT(\bx'|\by)$ which belong to the same bin as $\bx$.
Now,
\begin{eqnarray}
\bE\left\{N^s(\bx'|\bx,\by)\right\}&\exe&\left\{\begin{array}{ll}
\exp\{ns[\hat{H}_{\bx'\by}(X|Y)-R]\} & \hat{H}_{\bx'\by}(X|Y)>R\\
\exp\{n[\hat{H}_{\bx'\by}(X|Y)-R]\} & \hat{H}_{\bx'\by}(X|Y)\le
R\end{array}\right.\\
&=&\exp\{n(s[\hat{H}_{\bx'\by}(X|Y)-R]-(1-s)[R-\hat{H}_{\bx'\by}(X|Y)]_+)\},
\end{eqnarray}
Thus,
\begin{eqnarray}
&&\bE\left\{\left[\sum_{\bx'\ne\bx}
P(\bx',\by)\calI[f(\bx')=f(\bx)]\right]^s\right\}\\
&\exe&P^s(\by)\sum_{\calT(\bx'|\by)}P^s(\bx'|\by)\exp\{n(s[\hat{H}_{\bx'\by}(X|Y)-
R]-(1-s)[R-\hat{H}_{\bx'\by}(X|Y)]_+)\}\\
&=&P^s(\by)\sum_{\calT(\bx'|\by)}P^s(\bx'|\by)\exp\{n(s[\hat{H}_{\bx'\by}(X|Y)-
R]-(1-s)[R-\hat{H}_{\bx'\by}(X|Y)]_+)\}\\
&=&P^s(\by)\sum_{\calT(\bx'|\by)}\exp\{-n(s[D(\hat{P}_{\bx'|\by}\|P_{X|Y}|\hat{P}_{\by})+R]+
(1-s)[R-\hat{H}_{\bx'\by}(X|Y)]_+)\}\\
&\exe&P^s(\by)\exp\left\{-n\min_{P_{X'|Y}}(s[D(P_{X'|Y}\|P_{X|Y}|\hat{P}_{\by})+R]+
(1-s)[R-H(X'|Y)]_+)\right\}\\
&\dfn& P^s(\by)e^{-nL(\hat{P}_{\by},R,s)},
\end{eqnarray}
where $\hat{P}_{\bx'|\by}$ is the empirical conditional distribution of a
random variable $X'$ given $Y$ induced
by $(\bx',\by)$, and $D(P_{X'|Y}\|P_{X|Y}|P_Y)$ is defined as
\begin{equation}
D(P_{X'|Y}\|P_{X|Y}|P_Y)=\sum_yP_Y(y)\sum_{x}P_{X'|Y}(x|y)\log\frac{P_{X'|Y}(x|y)}{P_{X|Y}(x|y)}.
\end{equation}
Consequently,
\begin{eqnarray}
\overline{\mbox{Pr}\{\calE_1\}}&\le&
e^{nsT}\sum_{\bx,\by}P^{1-s}(\bx,\by)P^s(\by)e^{-nL(\hat{P}_{\by},R,s)}\\
&=&e^{nsT}\sum_{\by}P(\by)e^{-nL(\hat{P}_{\by},R,s)}\sum_{\bx}P^{1-s}(\bx|\by)\\
&=&e^{nsT}\sum_{\by}P(\by)e^{-nL(\hat{P}_{\by},R,s)}\prod_{i=1}^n\sum_{x\in\calX}P^{1-s}(x|y_i)\\
&\exe& e^{-nE_1'(R,T,s)}
\end{eqnarray}
where
\begin{equation}
E_1'(R,T,s)=\min_{P_Y'}\left[D(P_Y'\|P_Y)+L(P_Y',R,s)-
\sum_{y\in\calY}P_Y'(y)\ln\sum_{x\in\calX}P^{1-s}(x|y)\right]-sT.
\end{equation}
Finally,
\begin{equation}
E_1'(R,T)=\sup_{s\ge 0} E_1'(R,T,s).
\end{equation}

\section*{Appendix A}
\renewcommand{\theequation}{A.\arabic{equation}}
    \setcounter{equation}{0}

\noindent
{\bf Calculation of $L(R,s)$.}
Let $p_s=p^s/[p^s+(1-p)^s]$. Consider first the case $s\in[0,1]$, where
$p_s\ge p$. In this case, the minimizer $\delta^*$ that achieves $L(R,s)$ is
given by
\begin{equation}
\delta^*=\left\{\begin{array}{ll}
p & R< h(p)\\
h^{-1}(R) & h(p)\le R < h(p_s)\\
p_s & R\ge h(p_s)\end{array}\right.
\end{equation}
Here, for $R < h(p)$, the derivative of the objective function vanishes only
at $\delta=p >
h^{-1}(R)$, where the term $[R-h(\delta)]_+$ vanishes. On the other hand, for
$R\ge h(p_s)$, the derivative vanishes only at $\delta=p_s < h^{-1}(R)$, where
the term $[R-h(\delta)]_+$ is active. In the intermediate range, the
derivative jumps from a negative value to a positive value at
$\delta=h^{-1}(R)$ discontinuously, hence it is a minimum.
Thus, for $0\le s\le 1$, we have:
\begin{equation}
L(R,s)=\left\{\begin{array}{ll}
s[p\ln\frac{1-p}{p}+R-h(p)] & R<h(p)\\
sh^{-1}(R)\ln\frac{1-p}{p} & h(p)\le R < h(p_s)\\
sp_s\ln\frac{1-p}{p}+R-h(p_s) & R\ge h(p_s)\end{array}\right.
\end{equation}
For $s > 1$, $p_s< p$. and so $h(p_s)< h(p)$. Here, for $R < h(p_s)$, which
means also $R < h(p)$, the derivative vanishes only at $\delta=p> h^{-1}(R)$.
On the
other hand, for $R> h(p)>h(p_s)$,
the derivative vanishes only at $\delta=p_s< h^{-1}(R)$. In the intermediate
range, $h(p_s)\le R < h(p)$, the derivative vanishes both at $\delta=p$ and
$\delta=p_s$, so
the minimum is the smaller between the two. Namely, it is $\delta^*=p_s$ if
$$sp_s\ln\frac{1-p}{p}+s[R-h(p_s)]+(1-s)[R-h(p_s)]_+\le
sp\ln\frac{1-p}{p}+s[R-h(p)]+(1-s)[R-h(p)]_+$$
or equivalently,
$$sp_s\ln\frac{1-p}{p}+R-h(p_s)\le
sp\ln\frac{1-p}{p}+s[R-h(p)],$$
and it is $\delta=p^*$ otherwise. The choice between the two depends on $R$.
Let
\begin{equation}
R(s)=\frac{s(p_s-p)\ln[(1-p)/p]+sh(p_s)-h(p)}{s-1}
=-\frac{\ln[p^s+(1-p)^s]}{s-1}
\end{equation}
Then, for $s> 1$,
\begin{equation}
L(R,s)=\left\{\begin{array}{ll}
s[p\ln\frac{1-p}{p}+R-h(p)] & R<R(s)\\
sp_s\ln\frac{1-p}{p}+R-h(p_s) & R\ge R(s)\end{array}\right.
\end{equation}

\section*{Appendix B}
\renewcommand{\theequation}{B.\arabic{equation}}
    \setcounter{equation}{0}

\noindent
{\bf Calculations of Error Exponents for Very Weakly Correlated BSS's.}
For $p=1/2-\epsilon$, we have, to the second order in
$\epsilon$, $H(X|Y)=h(p)=h(1/2-\epsilon)=\ln 2 -2\epsilon^2$. Consider the
range of rates $\ln 2-2\epsilon^2 < R \le \ln 2$.
A second order Taylor series expansion of
$\gamma(t)\dfn-\ln[(1/2-\epsilon)^t+(1/2+\epsilon)^t]$
around $\epsilon=0$ (for fixed $t$) gives
\begin{equation}
\gamma(t)=(t-1)(\ln 2-2t\epsilon^2),
\end{equation}
and so,
\begin{eqnarray}
E_0(\rho,s)&=& \gamma(1-s)+\rho\gamma\left(\frac{s}{\rho}\right)\\
&=& -s[\ln 2-2(1-s)\epsilon^2]+(s-\rho)\left(\ln 2-
\frac{2s\epsilon^2}{\rho}\right)\\
&=&4s\epsilon^2-2s^2\left(1+\frac{1}{\rho}\right)\epsilon^2-\rho\ln 2.
\end{eqnarray}
Now,
\begin{equation}
E_1(R,T)=\max_{0\le s\le\rho\le 1}\left[s(4\epsilon^2-T)-\rho(\ln 2
-R)-2s^2\epsilon^2\left(1+\frac{1}{\rho}\right)\right].
\end{equation}
We will find it convenient to present $R=\ln 2-2\theta^2\epsilon^2$, where
$\theta\in[0,1]$, and so, from here on, the rate is parametrized by $\theta$.
The maximization over $\rho\ge s$, for a given $s$, is readily
found to give
\begin{equation}
\rho_s^*=s|\epsilon|\sqrt{\frac{2}{\ln 2-R}}=\frac{s}{\theta} \ge s,
\end{equation}
On substituting $\rho=\rho_s^*$, we get
\begin{eqnarray}
E_1(R,T)&\le&\max_{0\le s\le 1} [E_0(\rho_s^*,s)+\rho_s^*R-sT]\\
&=&\max_{0\le s\le 1}\left[s(4\epsilon^2-T)-s|\epsilon|\sqrt{2(\ln
2-R)}-2s^2\epsilon^2-2s|\epsilon|\sqrt{\frac{\ln 2-R}{2}}\right]\\
&=&\max_{0\le s\le 1}\{s[4\epsilon^2-T-2|\epsilon|\sqrt{2(\ln
2-R)}]-2s^2\epsilon^2\}\\
&=&\max_{0\le s\le 1}\{s[4\epsilon^2(1-\theta)-T
]-2s^2\epsilon^2\}
\end{eqnarray}
where the inequality is because when we maximized over $\rho$, we have ignored
the constraint $\rho\le 1$. Next, let $T = -\tau\epsilon^2$ for $\tau> 4$, then $s^*=1$
and so,
\begin{equation}
E_1(R,T)\le
4\epsilon^2(1-\theta)+\tau\epsilon^2-2\epsilon^2=2\epsilon^2(1-2\theta)+\tau\epsilon^2\le
(\tau+2)\epsilon^2.
\end{equation}
On the other hand,
\begin{eqnarray}
E_1'(R,T)&\ge&\sup_{s\ge 1}[R-sT+\gamma(s)+\gamma(1-s)]\\
&=&\sup_{s\ge 1}[s(4\epsilon^2-T)-4s^2\epsilon^2]+R-\ln 2\\
&=&\sup_{s\ge 1}[s(4\epsilon^2-T)-4s^2\epsilon^2]-2\theta^2\epsilon^2\\
&=&\frac{(4\epsilon^2-T)^2}{16\epsilon^2}-2\theta^2\epsilon^2\\
&\ge&\frac{[(\tau+4)\epsilon^2]^2}{16\epsilon^2}-2\epsilon^2\\
&=&\left[\frac{\tau(\tau+8)}{16}-1\right]\epsilon^2.
\end{eqnarray}

\end{document}